\documentclass[aps,twocolumn]{revtex4-1} 
\usepackage{hyperref} 
\usepackage[bottom]{footmisc}
\usepackage{graphicx}
\usepackage{amsmath}
\usepackage{float}
\usepackage{subcaption}
\usepackage{braket}
\usepackage{xcolor}
\begin{document}
\title{Understanding Basic Concepts of Topological Insulators Through Su-Schrieffer-Heeger (SSH) Model} 
\author{Navketan Batra and Goutam Sheet}
\affiliation{
Department of Physical Sciences, 
Indian Institute of Science Education and Research Mohali, 
Mohali, Punjab, India}

\begin{abstract}

Topological insulators are a new class of materials that have attracted significant attention in contemporary condensed matter physics. They are different from the regular insulators and they display novel quantum properties that also involve the idea of `topology', an area of mathematics.  Some of the fundamental ideas behind the topological insulators, particularly in low-dimensional condensed matter systems such as poly-acetylene chains, can be understood using a simple one-dimensional toy model popularly known as the Su-Schrieffer-Heeger model or the SSH model. This model can also be used as an introduction to the  topological insulators of higher dimensions. Here we give a concise description of the SSH model along with a brief review of the background physics and attempt to understand the ideas of topological invariants, edge states, and bulk-boundary correspondence using the model.

\end{abstract}
\maketitle

\section{Introduction}\label{intro}

Topology is a branch of mathematics that deals with different classes of geometries of objects. If you are allowed to bend and stretch an object but not tear it apart or join two regions together, then all the geometries that you will be able to make from the given object will be topologically equivalent. 
\begin{figure}[]
\centering
\includegraphics[scale=0.45]{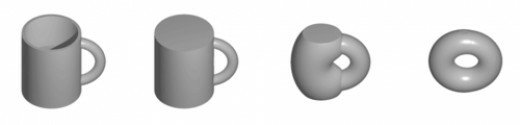}
\caption{Topological equivalence of a coffee mug and a doughnut: A coffee mug can be {\it smoothly} deformed into a doughnut without tearing apart or joining two regions together.}\label{fig:mug}
\end{figure}
A coffee mug is topologically equivalent to a doughnut, see FIG. \ref{fig:mug}. Similarly, a football is topologically equivalent to the page of the book you are probably reading this article in. Some of you might have encountered the Gauss-Bonnet's theorem in mathematics,
\begin{eqnarray}\label{gbth}
\frac{1}{4\pi}\int_M \kappa dA & = & 1-g
\end{eqnarray}
Here $\kappa$ is the Gauss curvature of the object, $M$, we are considering. What this statement means is that if we integrate an object's Gaussian curvature over the object's surface, divide it by $4\pi$, the result will be $(1-g)$ where $g$ is an {\it integer} representing the number of holes in the object (or a manifold) $M$. Making small deformations on the surface of the manifold will of course change the curvature of the object {\it locally} but when we integrate the `changed' curvature onto the deformed surface the right side of the equation still remains the same! Therefore, the number $g$ is known as a `topological invariant' meaning it does not change by small deformations on the surface of the manifold. It has recently been discovered that in condensed matter physics, it is possible to realize certain phases of matter that display unique physical properties that can be described by the existence of quantities analogus to the topological invariant $g$. One class of such materials is topological insulators. In this article we shall discuss this connection between the ideas of topology and topological invariants with condensed matter physics in the light of a beautiful toy model called the SSH model \cite{ssh1,ssh2}. However, before introducing the SSH model, we give a brief review of the background physics and discuss band insulators in section \ref{sec:bandinsulators}. In section \ref{sec:tightbinding}, we understand how metals and insulators are modelled using quantum mechanics by writing a {\it tight-binding Hamiltonian}. Then, in section \ref{sec:peierls}, we digress a little to explore Peierls instability, a special case of {\it charge density waves}. We then introduce the concept of Berry phase in section \ref{secbphase} which forms the basis of the theory of topological Insulators. Finally, in section \ref{sec:sshmodel}, we introduce the SSH model and do a detailed analysis before giving concluding remarks in section \ref{sec:conclusions}.

\section{The band insulators}\label{sec:bandinsulators}
To understand the ideas of topological Insulators, we first need to understand the Band Theory. Using a simple analogy with the particle-in-a-potential-well problem, let us explore how energy bands are formed in a solid. Consider a particle in a finite potential well from quantum mechanics - with certain approximations it can be seen as a model atom (hydrogen-like), where the potential well is created by the nucleus and the electron is a particle under that potential \cite{simon}. The one-dimensional version of this problem is solved in a standard undergraduate level quantum mechanics course \cite{rshankar}. The energy of such a system is discrete and can be computed to be $E_n = \frac{n^2\hbar^2 \pi^2}{2mL^2}$, where $L$ is the length of the box and $n$ is a level index (a non-zero +ve integer). Recall that $n$ = 1 corresponds the ground state energy. Now If we bring a second such finite well system close to the first, the full system can be treated as a model molecule with two electrons under the influence of the potential due to two nuclei. As per Pauli's exclusion principle, these two electrons cannot be accommodated in a single ground state unless their spins are anti-aligned. In fact, for such a molecule, the states corresponding to isolated particles overlap and we obtain two new energy eigenstates one having {\it lower} energy than the ground state energy of a single atom (the so-called bonding state for which the wavefunction is symmetric) and the other one having higher energy (the anti-bonding state for which the wavefunction is antisymmetric). In a hydrogen-like molecule, therefore, it makes sense for both the electrons to anti-align and occupy the bonding energy (symmetric) state to lower the energy of the whole system. Therefore we see that by bringing two atoms close to each other can in principle lower the energy of the whole system by the formation of bonding and anti-bonding states. Note that here we have ignored the Coulomb interaction between two charged electrons.

Now instead of just two atoms, if we take many such atoms to form a solid, there are several of these bonding and anti-bonding states. In case of a real macroscopic solid where the number of atoms is of the order of $10^{23}$, these states have to be so closely spaced that the discreteness of their energies can be ignored and instead we will have bands where such states exist. In a more complex solid, where each atom contributes several electrons, many distinct energy bands can form. Two such energy bands can also be separated from each other giving rise to a range of energies where no electronic states are present. Such an energy range is called a band gap in solid state physics and electronics \cite{kittel, mermin-ashcroft}. If the maximum energy up to which all quantum states are filled at zero temperature $T=0$, which is known as the Fermi energy of a solid, falls within such a band gap, the solid behaves like an insulator. Such insulators are often termed as the ``band insulators". If the Fermi energy is within a band, the solid behaves like a metal.  Notice, this theory, known as the band theory, neglects electronic interactions due to which it may fail for certain materials. 


\section{Tight binding model}\label{sec:tightbinding}
Imagine again building up a one-dimensional crystal lattice by arranging hydrogen-like atoms together along a straight line in an ordered fashion. As we discussed before, when atoms are far apart, they can be treated separately as there is no overlap between their respective eigenstates. Once they come close enough, their wave functions will no longer be orthogonal, they mix up and as a consequence, form bonding and anti-bonding states. Now with many atoms together, forming a crystalline lattice, the overlaps help the electron delocalise on the lattice by tunneling from one atom to the other. This tunneling potential is given by the overlap (inner product integral), denoted by $-t$, between the two states at different sites \cite{simon}.

\begin{figure}[]
\centering
\includegraphics[scale=0.243]{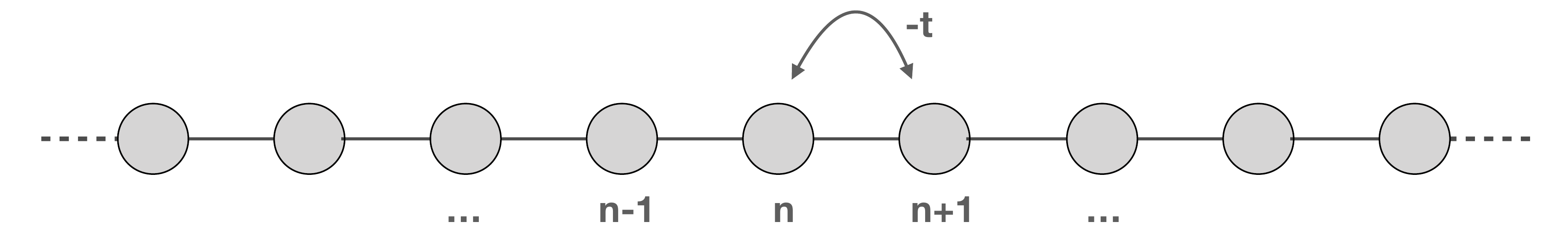}
\caption{Pictorial representation of a one dimensional tight-binding Hamiltonian.}
\end{figure}
One can now write down a Hamiltonian describing a particle's motion on a lattice structure - say, a one dimensional ring with $N$ such lattice points: 
\begin{eqnarray}
H & = & -\sum_{ij}t_{ij}\ket{i}\bra{j} 
\end{eqnarray} 
Here $i,j$ represent the lattice site and $-t_{ij}$ is the overlap integral between the $i$th and $j$th site. The sum over $i,j$ runs independently over {\it all} $i,j$ pairs. Note since going to-and-fro between two lattice points is equivalent, we impose $t_{ij} = t_{ji}$ (this condition makes the Hamiltonian Hermitian)  and $t_{ii} = 0$ for all site $i$. From quantum mechanics, we recall that the translation operator translates site $i$ to $i+1$ which we represent here as $T=\sum_i \ket{i+1}\bra{i}$. Now if we assume that the overlap integral only depends on the distance between the two sites $t_{i,i+n}\equiv t_n$ then we see that the tight binding Hamiltonian defined above is just a function of the translation operator and can thus be written as $H=-\sum_n t_{n} T^n$. Since now $[H,T]=0$, we can diagonalise the Hamiltonian in the the eigen-basis of the translation operator which we know are the plane wave states defined as 
\begin{eqnarray}
\ket{k} &= &\frac{1}{\sqrt{N}}\sum_j e^{ikj}\ket{j}
\end{eqnarray}
Action of $T^m$ on $\ket{k}$ gives $T^m\ket{k}=e^{-ikma}\ket{k}$. Therefore the eigenvalue spectrum for our tight binding Hamiltonian is given by 
\begin{eqnarray}
& H \ket{k} & = ~ \epsilon(k)\ket{k} \nonumber \\ 
& \epsilon (k) & = ~ -\sum_n t_n e^{-ikna}
\end{eqnarray}
For a case when we only allow the nearest neighbour hopping i.e. $t_1=t\ne0$ and all others are zero. Then, we obtain $\epsilon(k)=-2t\cos{ka}$. 

\begin{figure}[]
\centering
\includegraphics[scale=0.3]{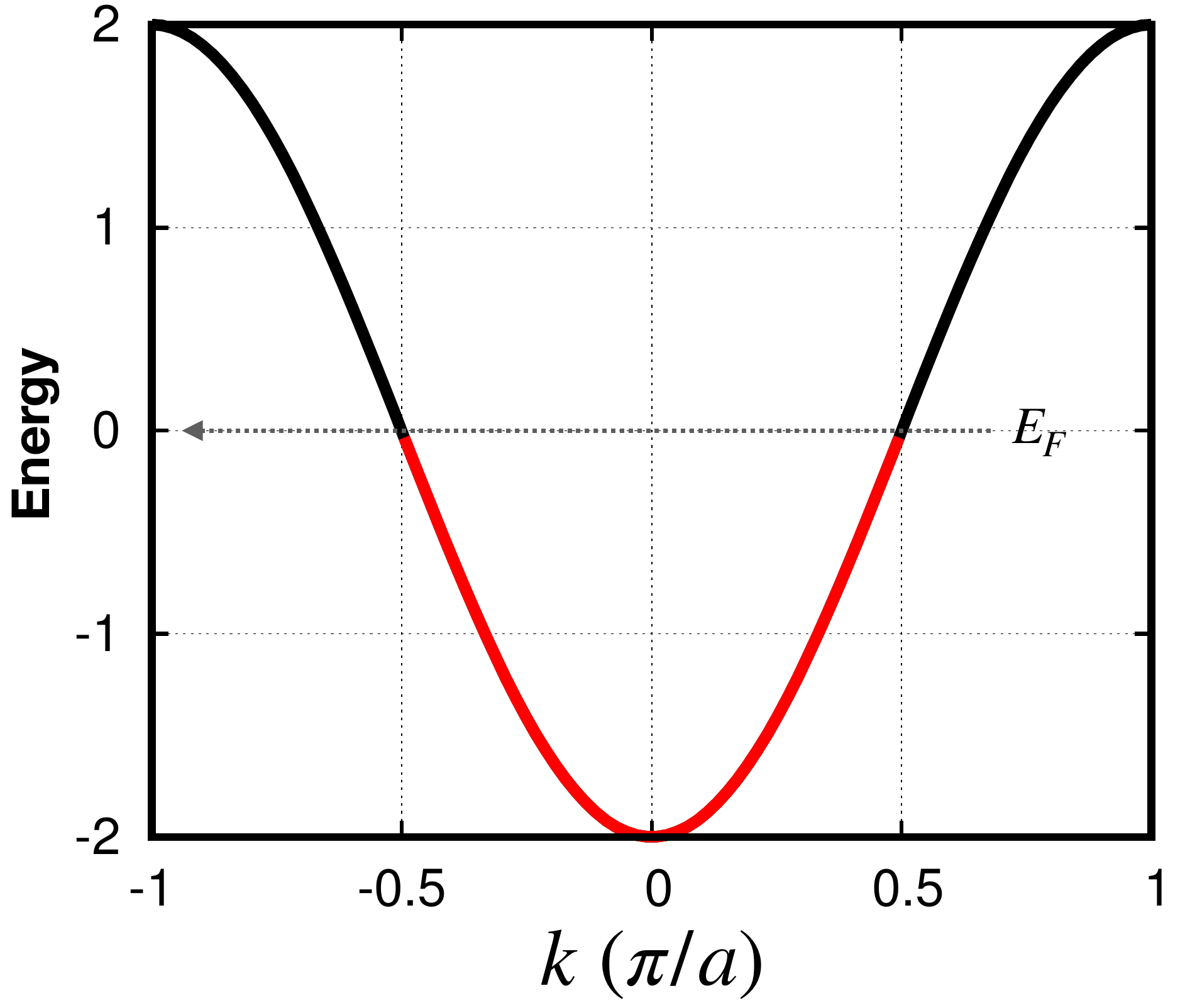}
\caption{Energy dispersion obtained from a 1D nearest neighbour tight binding Hamiltonian. The red portion represent the filled states.}
\label{tightbindingdispersion}
\end{figure}

This is the dispersion that is followed by a single electron hopping on a lattice made from hydrogen-like atoms. It turns out that this method, when applied to the many-electron problem by neglecting the $e^--e^-$ interaction, is not a bad approximation to describe many physical properties like metallic character. All the allowed $k$ falling in the line from $-\pi/a$ to $+\pi/a$ constitute a Brillouin zone. The edges at $k = \pm \pi/a$ are called the Brillouin zone edges which have interesting features when weak interactions are introduced - we will come back to this in the next section.
If we stick to non-interacting particle description and assume every atom has a single loosely bound electron that hops around, then the band formed by the dispersion relation is half filled (red region in FIG. \ref{tightbindingdispersion}). This is because every atomic state contributes one electron whereas its maximum occupancy is restricted to two (due to Pauli's principle). Notice that in condensed matter, energies are always measured from the Fermi energy so it is convenient to set Fermi energy to zero. Since there are states available at arbitrary small energies above Fermi energy, this system thus describes a metal otherwise, an insulator as we will see in the next section.

\section{Peierls instability}\label{sec:peierls}
In the previous problem, we studied a one-dimensional non interacting tight-binding model that we saw described a metal. When we have drawn the $E$ vs. $k$ relation in FIG. \ref{tightbindingdispersion}, we assumed the existence of background lattice which give rise to a periodic potential that the electrons can sense. Now consider several free electron dispersions (which is just parabolic $E=\frac{\hbar^2 k^2}{2m}$) separated from each other.
\begin{figure}[]
\centering
\includegraphics[scale=0.23]{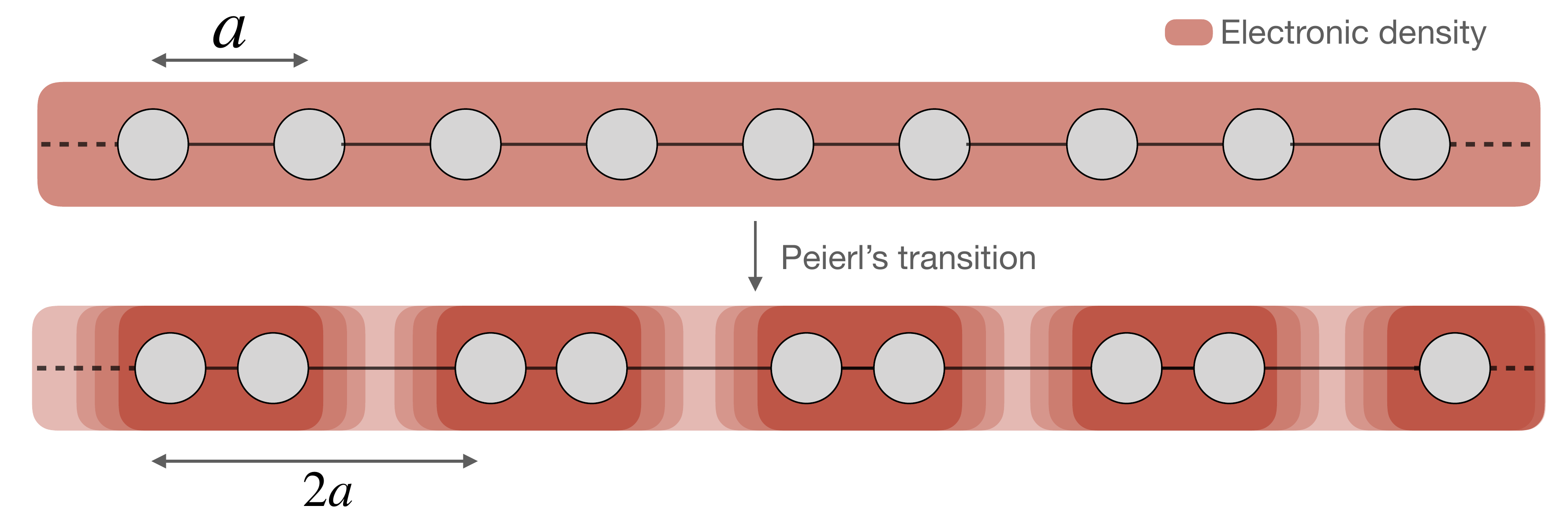}
\caption{A 1D lattice, lattice constant $a$, undergoes a Peierls transition and distorts the lattice such the the lattice constant becomes $2a$. The brown region shows the electronic density.}
\label{peierlsfig}
\end{figure} 
The point where two dispersions cross each other is a point of degeneracy which will be lifted once weak periodic potential due to the lattice is included. For the similar reason energy gaps arise in their dispersion relation at the Brillouin zone boundary when weak periodic potential is considered. If we put just one electron per lattice site then the states from $-\pi/2a$ to $+\pi/2a$ will be occupied without a band gap between the occupied and the unoccupied states. Hence, the system will still behave like a metal. However, in 1930s, Rudolf Peierls proved that a one-dimensional equally spaced chain with one electron per lattice site is unstable and such a system is prone to distortion. This can be qualitatively understood in the following way.


\begin{figure}[]
\centering
\includegraphics[scale=0.37]{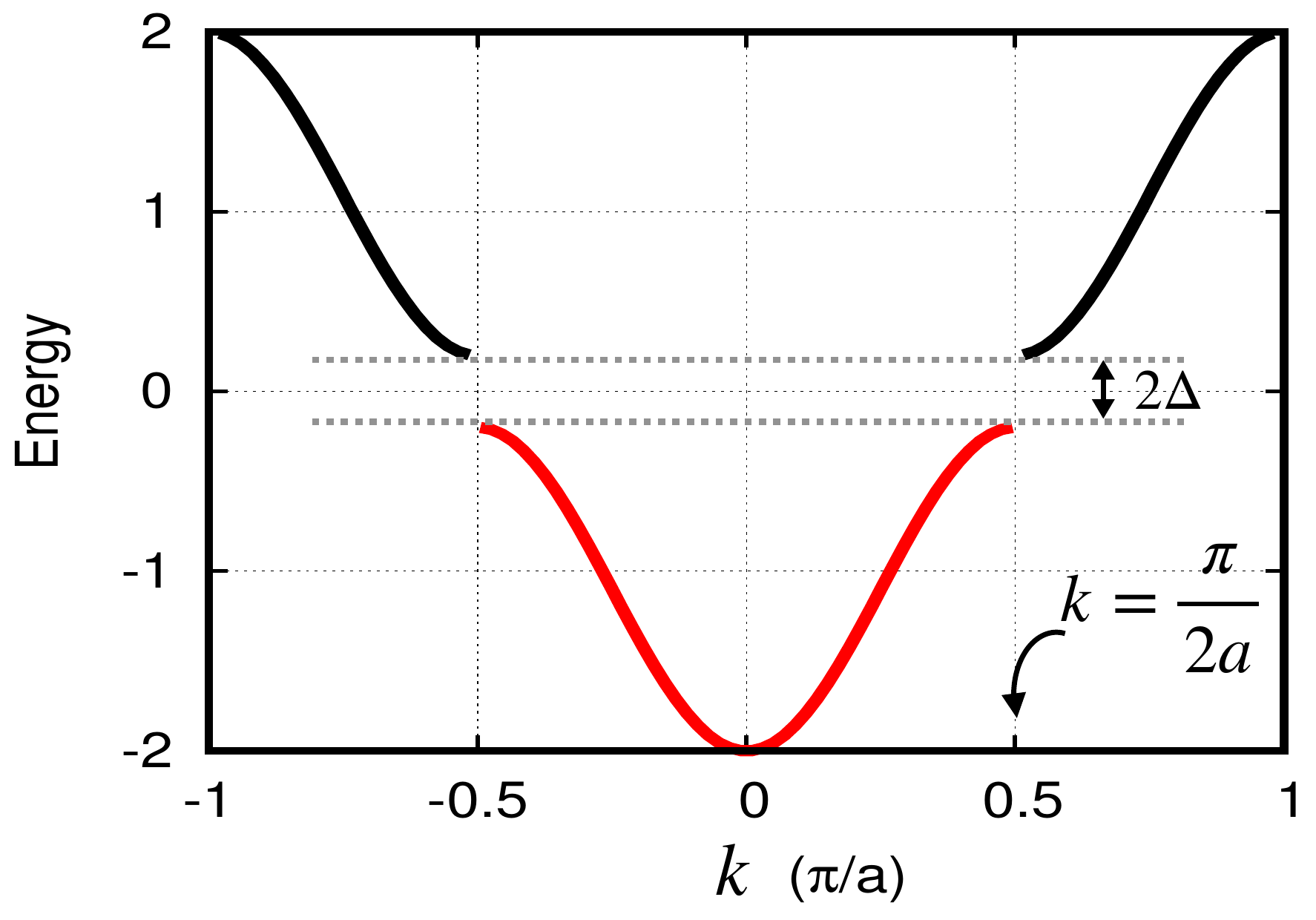}
\caption{Due to the distortion in the lattice, the lattice constant is now $2a$ instead of $a$ therefore the Brillouin zone boundary is at $k=\pm \frac{\pi}{2a}$}
\label{pinsulator}
\end{figure}
In FIG. \ref{peierlsfig}, the brown region shows the electronic density due to the electrons. In the normal case, where the lattice spacing is $a$, the electronic density is uniform. But if we consider a case where every atom distorts itself such as the resulting lattice has two sublattice sites in a unit cell and the inter-unit cell distance is $2a$. The electronic density will become periodically modulated. This is known as a \textbf{charge density wave} \cite{thorne}. The final energy of the distorted lattice atoms alone will, of course, be higher than the original undistorted case. In fact, estimating this energy per unit length in the elastic approximation gives
\begin{eqnarray}
\Delta E_{\text{lattice}} = \frac{1}{4}K\delta^2
\end{eqnarray}
which is quadratic in the distortion $\delta$. $K$ is the elastic constant determined by the material.
However, if we calculate the electronic energy per unit length of the new charge configuration due to distortion, in the low distortion limit we get,
\begin{eqnarray}
\Delta E_{\text{electronic}} \approx \frac{A^2\delta^2}{\pi} \ln{|\delta|}
\end{eqnarray}
the total energy per unit length, $\Delta E_{\text{lattice}} + \Delta E_{\text{electronic}}$ is extremized as,
\begin{eqnarray}
\frac{d}{d\delta} [ \Delta E_{\text{total}} ] = \frac{1}{2} K \delta + \frac{2A^2}{\pi}\delta \ln{|\delta|} + \frac{A^2}{\pi}\delta = 0
\end{eqnarray}
The solution $\delta = 0$ corresponds to a maxima which can be checked explicitly by calculating the second derivative. The minima corresponds to  
\begin{eqnarray}
\delta_{\text{min}} = \pm e^{-\frac{1}{2}} e^{-\frac{K}{4\pi}}
\end{eqnarray}
It can also be checked, by plugging in $\delta_{\text{min}}$ into $\Delta E_{\text{total}}$, that total change in energy is negative! This means that the system gains energy by distorting itself by $\delta_{\text{min}}$. Note that is was only possible due to electronic energy that compensated the higher-than-original energy contributed by the lattice due to distortion. This is called \textbf{Peierls instability}.

As illustrated in FIG. \ref{pinsulator}, now it is easy to notice that for the distorted lattice, the zone-edges are located at $k = \pm \pi/2a$ instead of at $k = \pm \pi/a$ since now the periodicity of the Brillouin zone is doubled. Doubling the period hence introduces new band gaps in the Brillouin zone and therefore system becomes lower in energy corresponding to the states in the vicinity of the new gaps. This transition makes the band completely filled and a band gap (shown as $2\Delta$ in FIG. \ref{pinsulator}) opens up separating the empty states from the filled states at the new edges of the Brillouin zone. This makes the system a band insulator.

\section{Berry Phase}\label{secbphase}
Let us digress a little and look at a very interesting concept in quantum mechanics that will become necessary at a later stage - the concept of Berry Phase. Consider a general Hamiltonian $H({\bf R})$, which is a function of several parameters represented as a vector ${\bf R} = (R_1,R_2,..)$. At any instant, for a fixed ${\bf R}$, the solutions can be obtained by using the time-independent Schroedinger equation as,
\begin{eqnarray}
H({\bf R}) \ket{n(\bf R)} = E_n({\bf R}) \ket{n(\bf R)}
\end{eqnarray} 
Now as ${\bf R}$ changes in the parameter space (starting from ${\bf R}(t=0)$) along some path $\mathcal{C}$, we are interested in knowing how the state changes when the system is initially prepared in the state $\ket{n({\bf R}(t=0))}$. There is a very useful theorem known as the {\it Adiabatic theorem} that states - for a {\it slowly} varying Hamiltonian, a system initially in the eigenstate will always remain in its {\it instantaneous} eigenstate at any later time. From this theorem we have got half the answer to our question. We now know starting from $\ket{n({\bf R}(0))}$, the system will evolve to $\ket{n({\bf R}(t))}$ which is the instantaneous eigenstate of $H({\bf R}(t))$ with slowly varying ${\bf R}$. But what about the phase? We can in general write the evolved state at $t$ as $\ket{\psi(t)} = e^{-i\theta(t)}\ket{n({\bf R}(t))}$. This state will follow the Schroedinger equation,
\begin{eqnarray}
H({\bf R(t)})\ket{\psi(t)} = i\hbar \frac{d}{dt}\ket{\psi(t)}
\end{eqnarray} 
which translates into the differential equation,
\begin{eqnarray}
E_n({\bf R}(t))\ket{n({\bf R}(t))} & = & \hbar \left( \frac{d}{dt} \theta(t) \right) \ket{n({\bf R}(t))} \nonumber \\ 
& & + i\hbar \frac{d}{dt} \ket{n({\bf R}(t))}
\end{eqnarray}
Taking the scalar product with $\bra{n({\bf R}(t))}$ and assuming the state is normalised, we get,
\begin{eqnarray}
E_n({\bf R}(t))- i\hbar \bra{n({\bf R}(t))} \frac{d}{dt} \ket{n({\bf R}(t))} = \hbar \left( \frac{d}{dt} \theta(t) \right) \nonumber
\end{eqnarray}
\begin{eqnarray}
 \theta(t) &=& \frac{1}{\hbar} \int^t_0 E_n({\bf R}(t'))dt' \nonumber \\
&&- i \int^t_0  \bra{n({\bf R}(t'))} \frac{d}{dt'} \ket{n({\bf R}(t'))} dt'
\end{eqnarray}
The first term of the phase is just the conventional dynamical phase that arises due to time evolution of the Hamiltonian. The negative of the second term is what is known as the Berry Phase $\gamma_n$,
\begin{eqnarray}
\gamma_n = i \int^t_0  \bra{n({\bf R}(t'))} \frac{d}{dt'} \ket{n({\bf R}(t'))} dt'
\end{eqnarray} 
This term arises because the states at $t$ and $t+dt$ are not `identical' and a phase is picked up that depends on the trajectory in the parameter space. From the previous expression,
\begin{eqnarray}\label{bphase}
\gamma_n &= & i \int^t_0  \bra{n({\bf R}(t'))} \nabla_{\bf R} \ket{n({\bf R}(t'))} \frac{d{\bf R}}{dt'} dt'\\
 &=& i \int^{{\bf R}_t}_{{\bf R}_0}  \bra{n({\bf R})} \nabla_{\bf R} \ket{n({\bf R})} d{\bf R}  \\
 &=&  \int^{{\bf R}_t}_{{\bf R}_0}  {\bf A}_n({\bf R}) d{\bf R}
\end{eqnarray}
where we define Berry potential as,
\begin{eqnarray}
{\bf A}_n({\bf R}) = i \bra{n({\bf R})} \nabla_{\bf R} \ket{n({\bf R})} 
\end{eqnarray}
From quantum mechanics, we know that by multiplying the states by an overall {\it global} phase factor,
\begin{eqnarray}
\ket{n({\bf R})} \rightarrow \ket{\tilde{n}({\bf R})}  = e^{i\chi({\bf R})} \ket{n({\bf R})} 
\end{eqnarray} 
the dynamics of the system does not change - this is known as {\it gauge invariance}. However, we see that the Berry Potential ${\bf A}_n({\bf R})$, is not a gauge invariant quantity. Under gauge transformation it changes as
\begin{eqnarray}
{\bf A}_n({\bf R}) \rightarrow {\bf A}_n({\bf R}) - \frac{\partial}{\partial {\bf R}} \chi({\bf R})
\end{eqnarray}
Cosequently, Berry phase will change by, $-\int_{\mathcal{C}} \frac{\partial}{\partial {\bf R}} \chi({\bf R}) d{\bf R} = \chi({\bf R_0}) - \chi({\bf R_t})$. Therefore, We can have Berry phase to be gauge invariant as long as the path $\mathcal{C}$ is closed. This quantity shows up in many areas of physics, one classic example is the Aharonov-Bohm effect \cite{ABeffect}. 

\section{The SSH Model}\label{sec:sshmodel}
\begin{figure}[]
\centering
\includegraphics[scale=0.235]{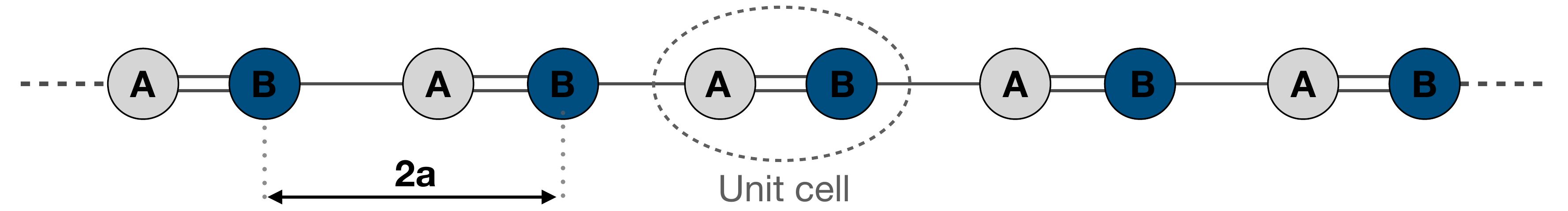}  
\caption{Visual representation of the SSH Model}
\label{sshfig}
\end{figure}

We now move onto a slightly more complicated version of the previous Hamiltonian where instead of a single site unit cell we have a two-site unit cell, as shown in FIG. \ref{sshfig}, as if the lattice has been distorted due to Peierls instability. The Su-Schrieffer-Heeger (SSH) model is a tight-binding model that describes a single spin-less electron on a two site unit cell 1D lattice. The two sites in a unit cell are labelled as $A$ and $B$. From here we will set the lattice constant, $2a=1$.

We will work with one electron per unit cell that corresponds to a half-filled lattice because we are considering spin-less electrons. Thus the only degree of freedom the electrons have is that they can hop around from one site to the other - let us call the hopping potentials as $v$ (for hopping within the unit cell) and $w$ (for hopping connecting neighbouring unit cells), and write the tight binding Hamiltonian as
\begin{eqnarray}
H & = & v\sum_{n=1}^{N} (\ket{n,B}\bra{n,A}+\text{h.c.}) \nonumber \\
&& +w\sum_{n=1}^{N}(\ket{n+1,A}\bra{n,B}+\text{h.c})
\end{eqnarray}
Here `h.c.' denotes the hermitian conjugate of the term before it. The basis we have used to define the Hamiltonian can also be written as $\ket{n,A(\text{or }B)}=\ket{n}\otimes\ket{A(\text{or }B)}$. This representation tells us that the full Hilbert space is made of two parts - $\mathcal{H}_{\text{external}}\otimes\mathcal{H}_{\text{internal}}$. The internal Hilbert space is due to the two sub-lattice sites ($A$ and $B$) in the unit cell and the external is due to the repetition of the unit cells $N$ times. 

\subsection{The band Hamiltonian}

\begin{figure}[]
\centering
\includegraphics[scale=0.34]{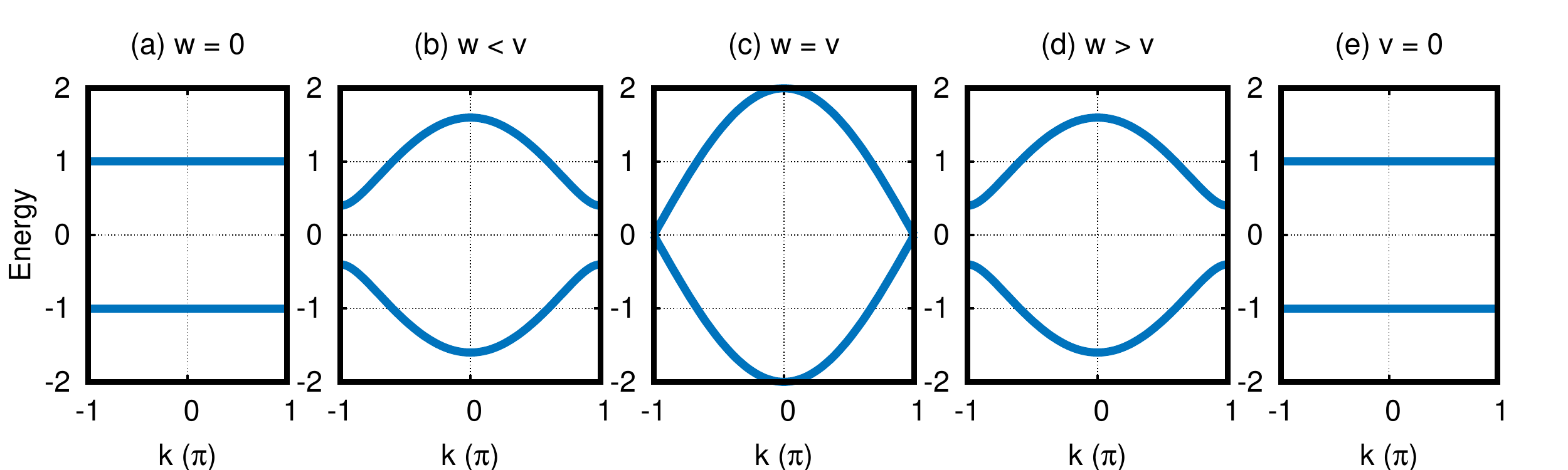}
\caption{Dispersion relation obtained from bulk Hamiltonian plotted for different parameter cases}
\label{sshdispersion}
\end{figure}

The two-site unit cell adds another band to the band structure we explored previously. To see this, we need to solve for its dispersion relation by only focusing on the bulk part of the chain, that is to say, assuming the chain forms a loop (periodic boundary). Because the system is translationally invariant, it allows us to perform Fourier transform in the external Hilbert space similar to what we did earlier. 
\begin{eqnarray}
\ket{n}\otimes\ket{A(\text{or }B)} = \frac{1}{\sqrt{N}} \sum_k e^{ikn} \ket{k}\otimes \ket{A(\text{or }B)}
\end{eqnarray}
Plugging this back into the Hamiltonian (with periodic boundaries), we get 
\begin{eqnarray}
H&=&\sum_k (v+e^{-ik}w)\ket{k,B}\bra{k,A}\nonumber \\
&& + (v+e^{ik}w)\ket{k,A}\bra{k,B}
\end{eqnarray} 
The external and internal parts can be separated and this can further be reduced to 
\begin{eqnarray}
H&=&\ket{k}\bra{k}\otimes\big[ (v+e^{ik}w) \ket{A}\bra{B}\nonumber \\
&&+ (v+e^{-ik}w) \ket{B}\bra{A} \big]
\end{eqnarray}
In a compact form, $H=\sum_k \ket{k}H(k)\bra{k}$. 

\begin{eqnarray}
H(k)=
\begin{pmatrix}
0 & v+e^{ik}w \\
v+e^{-ik}w 	& 0
\end{pmatrix}
\end{eqnarray}

$H(k)$ is the band Hamiltonian that acts on the internal Hilbert space \cite{rshankar2}. With the help of Fourier transform we have block diagonalised the full Hamiltonian into $N$ such $2\times 2$ band Hamiltonians for each $k$. Now our task is simply to diagonalise this two dimensional matrix that will give us the bulk dispersion relation in terms of the energy eigenvalues:
\begin{eqnarray}
& E(k)=\pm \sqrt{v^2+w^2+2vw\cos(k)}
\end{eqnarray}
Likewise, the eigenstates (eigenvectors) look like,
\begin{eqnarray}
 \ket{\pm(k)} = \begin{pmatrix}
\pm e^{-i\phi(k)} \\
1
\end{pmatrix} \nonumber \\
\phi(k)=\tan^{-1} \left(\frac{w\sin{k}}{v+w\cos{k}}\right) 
\end{eqnarray}

Note that our Hamiltonian has two parameters: $v$ and $w$. It is clear that different choices of these parameters will lead to different dispersion relations. Let us try to see what these band structures look like and what can one conclude from the dispersion. Note that here we have considered the Fermi energy to be the zero energy level -- this is the origin in our energy scale relative to which all other energies are measured here.

The plots in FIG. \ref{sshdispersion} suggest that when there is staggering, $v\ne w$, in the Hamiltonian, the dispersion is gapped and hence a staggered case would be an insulator. Only in the case of $v=w$ the gap closes and there are states available at arbitrary low energies above the Fermi level, therefore $v=w$ case is a metal. 
 
\subsection{Beyond energy-band description} 
 
From the Bulk Hamiltonian, it seems like the problem is symmetric about the $v=w$ case. By symmetric we mean that $v > w$ case is exactly the same as $w > v$ case. But this is not quite right. The information obtained from the `eigenvalues' is not complete! We need to also look at the eigenvectors to get a complete information about these cases. Interestingly, the topological aspect of this problem is hidden in the eigenvectors.  

Since the band Hamiltonian is a two dimensional Hermitian matrix it can be written in a more insightful notation as,
\begin{eqnarray}
H(k) &= &\vec{h}(k)\cdot\vec{\sigma}
\end{eqnarray} 
using Pauli's matrices, $\{\sigma_i\}$, as the basis for a two dimensional Hermitian matrix \cite{sakurai}.
By comparing terms we get, 
\begin{eqnarray}
h_x(k) & = & v+w\cos{k} \nonumber \\
h_y(k) & = & w\sin{k} \nonumber \\
h_z(k) & = & 0
\end{eqnarray}
As we saw earlier, the eigenstates are parametrised by $\phi=\tan^{-1} (\frac{w\sin{k}}{v+w\cos{k}})=\tan^{-1}{\frac{h_y(k)}{h_x(k)}}$. Therefore, in the $h_x-h_y$ space, the direction of the vector $\vec{h}(k)$ denotes an eigenstate and the magnitude of the vector will give its eigenvalue. 

\begin{figure}[]
\centering
\includegraphics[scale=0.24]{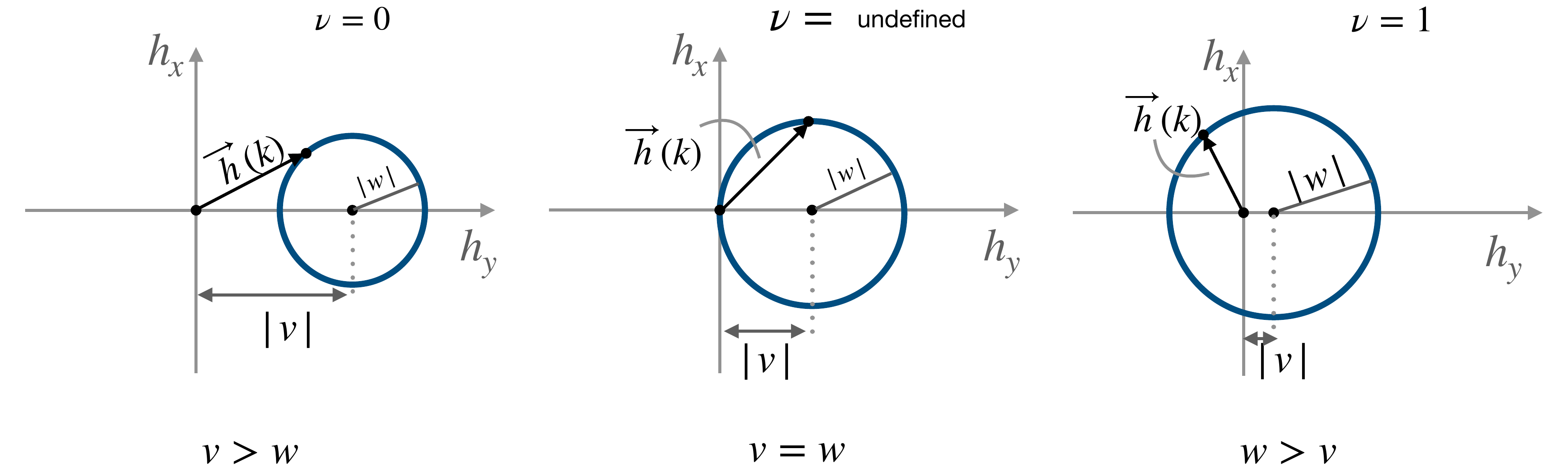}
\caption{Trajectories of $\vec{h}(k)$ over the Brillouin zone in the $h_x-h_y$ space for different parameter values}
\label{windingfig}
\end{figure}

A plot of the trajectory of $\vec{h}(k)$ over the first Brillouin zone \{$-\pi \rightarrow \pi$\} gives the full set of eigenstates. If we plot the trajectories for the two seemingly equivalent cases: $v<w$ and $v>w$ we observe, see FIG. \ref{windingfig}, that even though their dispersions are similar, these trajectories on the $h_x-h_y$ plane are very different!
The vector $\vec{h}(k)$ that denotes the eigenstates will necessarily form a closed loop as a consequence of the periodicity of the Brillouin zone. For one of the insulating cases ($v<w$), the vector $\vec{h}(k)$ winds about the origin and for the other case ($v>w$) it does not. Notice that the origin is the point where $\vec{h}(k)=0$ which is the gap-less condition. Therefore a loop plotted for the $v=w$ case would pass through the origin indicating a metallic state. We can call this as a {\it winding number} $\nu$ that tells whether the trajectory of $\vec{h}(k)$ winds about the origin or not. Winding number distinguishes the two seemingly equivalent cases.  

Notice that the bulk Hamiltonian we are looking at, $H(k)$, is also a function of some parameter, in this case $k$, which is known as the crystal momentum. Therefore, Berry potential for an energy band can be defined. The Brillouin zone is periodic and has a topology of a ring. If we consider the Brillouin zone as a parameter space, like we did in section \ref{secbphase}, and since it is periodic, integrating the Berry potential for the filled band over the Brillouin zone will correspond to tracing a closed loop in this parameter space and therefore will result in a gauge invariant berry phase. Let us now calculate the Berry phase \cite{rshankar2}, 
\begin{eqnarray}
A_{-}(k) = i\bra{-(k)} \frac{d}{dk} \ket{-(k)} = -\frac{1}{2} \frac{d\phi}{dk}
\end{eqnarray}
Now if we integrate this quantity in the Brillouin zone, and put appropriate factors, we get,
\begin{eqnarray}\label{nu}
\frac{-1}{\pi} \oint A_{-}(k) dk = \begin{cases}
    1 & \text{if $v < w$} \\
    0 & \text{if $v > w$} \\
    \text{undefined} & \text{if $v = w$}
  \end{cases}
\end{eqnarray}
Looking at this result quickly reminds us of the winding number $\nu$. In fact, eq. (\ref{nu}) is a proper way to calculate the winding number. Notice that it is undefined for the $v=w$ case since the notion of filled band itself is undefined. Notice that the form of eq. (\ref{nu}) is similar to eq. (\ref{gbth}) - let us try to see its analogy with the eq. (\ref{gbth}). The role of Gaussian Curvature is played by Berry potential; and the role of the Manifold we integrate it over is played by the Brillouin zone. Notice that the topological invariant $g$ which does not change with small deformations on the surface of the Manifold is analogous to the winding number. Small changes in the Berry Potential (or in the Hamiltonian via the parameters $v$ or $w$) will indeed change the Berry potential locally, but when it is integrated over the entire Brillouin zone, the winding number remains invariant! In other words, small changes to $h(k)$ will change the trajectory of the vector $\vec{h}(k)$ and the trajectory will no longer remain a perfect circle but the winding number will still be defined and remain unchanged. 

Now we have seen that the two seemingly equivalent insulating cases are `topologically' different which is concluded by exploiting the information of wavefnuctions in these two cases. But how do these different behaviours of the eigenvectors make the two insulating cases different physically? In other words, we are looking for a physical consequence of the winding number in the seemingly similar insulating cases. But before answering that in its full glory, let us note one physical consequence that can be drawn from this distinct behaviour. If we were to smoothly transform the Hamiltonian, in other words change any of the parameters $v$ or $w$, then since $h_z=0$, the only possible way to get from the insulating phase described by the case $v<w$ to the insulating phase described by the case $v>w$ (or the other way round), is via crossing the origin! What this means is that somewhere in the middle we have to get to the solution of eigenstates corresponding to a trajectory, $h(\vec{k})$, in the $h_x-h_y$ space, that passes through the origin in that space. Therefore, a smooth transition from one insulating phase to the other is not possible without closing the energy gap (or crossing the metallic phase) at least once. This is the hallmark of a \textbf{topological phase transition} \cite{kane}. An analogy can be drawn here with a 3D topological insulator (TI) like Bi$_2$Se$_3$. In a 3D TI, the bulk is an insulator which is topologically distinct (has different winding number) from the  vacuum (outside of the TI). Note that vacuum can be thought of like a band insulator with the particle (like electron) and the antiparticle (like hole) states separated by a ``band gap". In order to smoothly transition from the TI phase to the band insulator phase, one must cross a point (the interface between the two) where the band gap is zero. That gives rise to conducting surface states at the surface of 3D TIs. We will now explore these conducting states, known as {\it edge states} in 1D, that appear in the SSH model.  

\subsection{A finite SSH chain}
Till now we looked at the solutions of a 1D SSH chain with periodic boundaries (i.e. a ring). To answer the question posed in the last section on what physical consequence does the winding number has let us look at the version of the SSH chain with open boundaries. The problem is not trivial - since there is no translational invariance we cannot make use of the Fourier transformation to diagonalise the Hamiltonian. We, therefore, feed the Hamiltonian into our computers and explicitly solve the Hamiltonian for its eigenvalues and eigenfunctions. 

\begin{figure}[]
\centering
\includegraphics[scale=0.235]{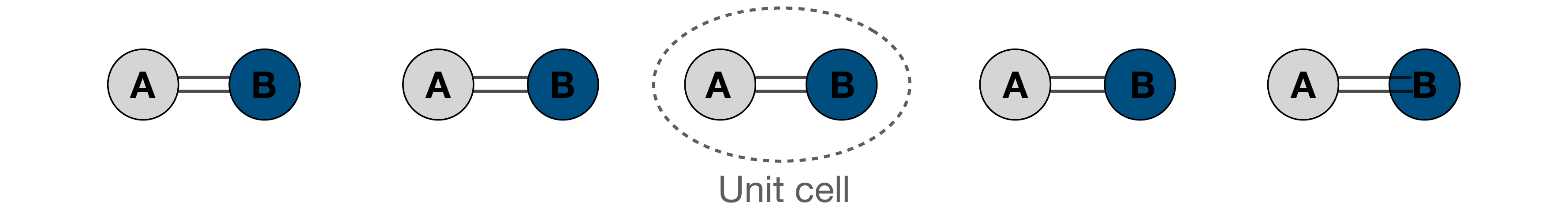}
\caption{Open SSH chain in the limit $w=0$ and $v\ne 0$}
\label{openssh1}
\end{figure}

But before this, let us first try to look at the extreme cases and guess what could be the solutions. There are two extreme cases - $w=0$,$v\ne 0$ and $v=0$,$w\ne 0$. These are called the dimerised limits as this would correspond to breaking the chain into dimers. 
The first case, $w=0$, for a chain with $N=10$ looks something like what is shown in FIG. \ref{openssh1}, 
This seems like a repetitive problem of a dimer which is, of course, easy to solve. We know the solutions should be the superposition of the $A$ site and $B$ site for each dimer. We can even guess that this should correspond to an insulating state as the chain is now broken and no particle can hop from one end of the chain to the other. 

Let us now look at the other extreme case of $v=0$. 
\begin{figure}[]
\centering
\includegraphics[scale=0.235]{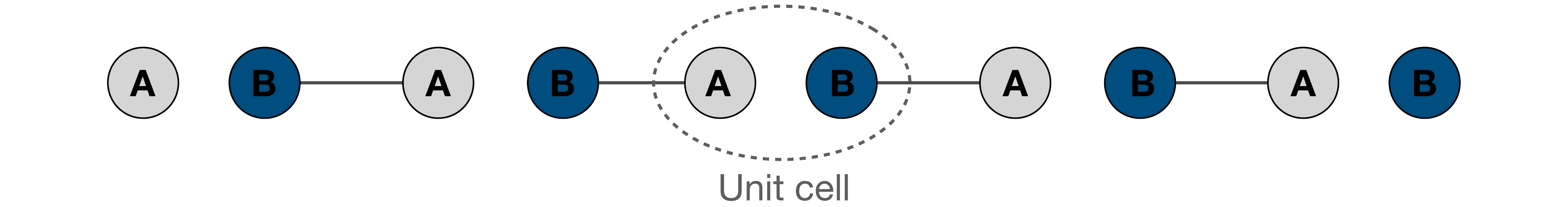}
\caption{Open SSH chain in the limit $v=0$ and $w\ne 0$}
\label{openssh2}
\end{figure}  
Similar to the previous case, we get a few dimers but since the chain has open boundaries, we now have two single sites at the end of the chain, see FIG. \ref{openssh2}. If these two sites, one at each end, carry an electron the energy should be zero because, in the SSH model, there is no energy contribution for an electron to be held fixed at one site. So we should expect two zero energy states in the system localised at the edges of the chain. Again going by the same argument as in the previous case we expect this case to describe an insulator as well. 

Let us now solve the Hamiltonian for an open SSH chain and try to see what are the solutions for different parameter values. 
\begin{figure}[]
\begin{subfigure}{.3\textwidth}
  \centering
  \includegraphics[scale=0.60]{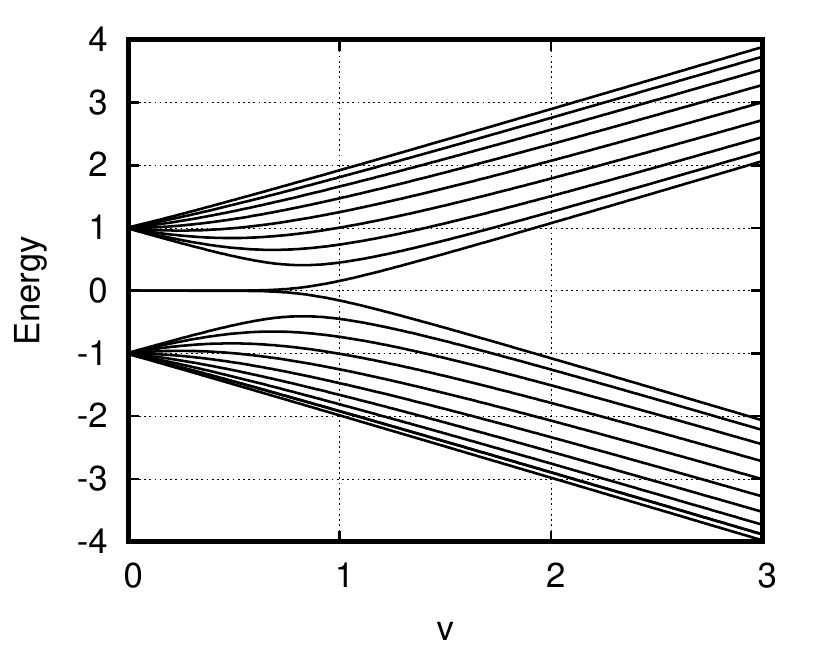}
  \caption{$w=1$}
  \label{fig:sfig1}
\end{subfigure} 
\begin{subfigure}{.3\textwidth}
  \centering
  \includegraphics[scale=0.60]{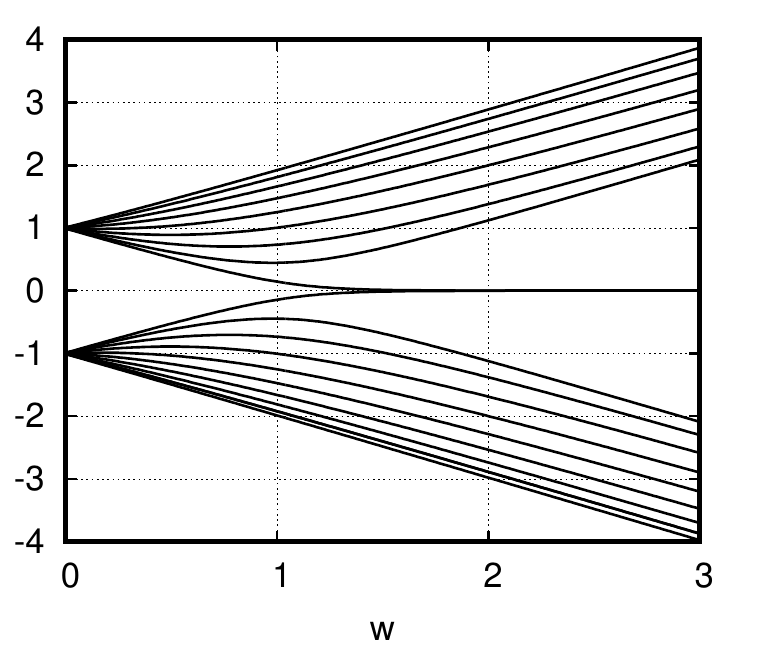}
  \caption{$v=1$}
  \label{fig:sfig2}
\end{subfigure}
\caption{Eigenvalue spectrum as a function of model parameter $v$ (case (a)) when $w$ is held fixed at $w=1$ and $w$ (case(b)) when $v$ is held fixed at $v=1$.}
\label{specssh}
\end{figure}
The plots in FIG. \ref{specssh} show the energy spectrum of the eigenstates as we vary the parameter (a) $v$ and (b) $w$. For the dimerised case, our expected solutions match perfectly well and we indeed see zero energy states in the case where $w=0$ and no zero energy states in the other $v=0$ case. One interesting feature we see here is that the zero energy states not only exist in the extreme limit but even for non-zero $v$, although the zero energy states are not quite zero energy states but are very close to zero. 
Here again, our Fermi energy lies at zero so we see that the spectrum is gapped in the case $w=0$. However, due to the presence of zero energy states in the $v=0$ case, we cannot be quite sure about its insulating behaviour. But intuitively a dimerised chain should not be conducting and also the zero energy states should be localised at the boundary. Therefore, let us look at the corresponding wavefunctions of these energy eigenstates \cite{asboth}.
\begin{figure}[]
\centering
\includegraphics[scale=0.35]{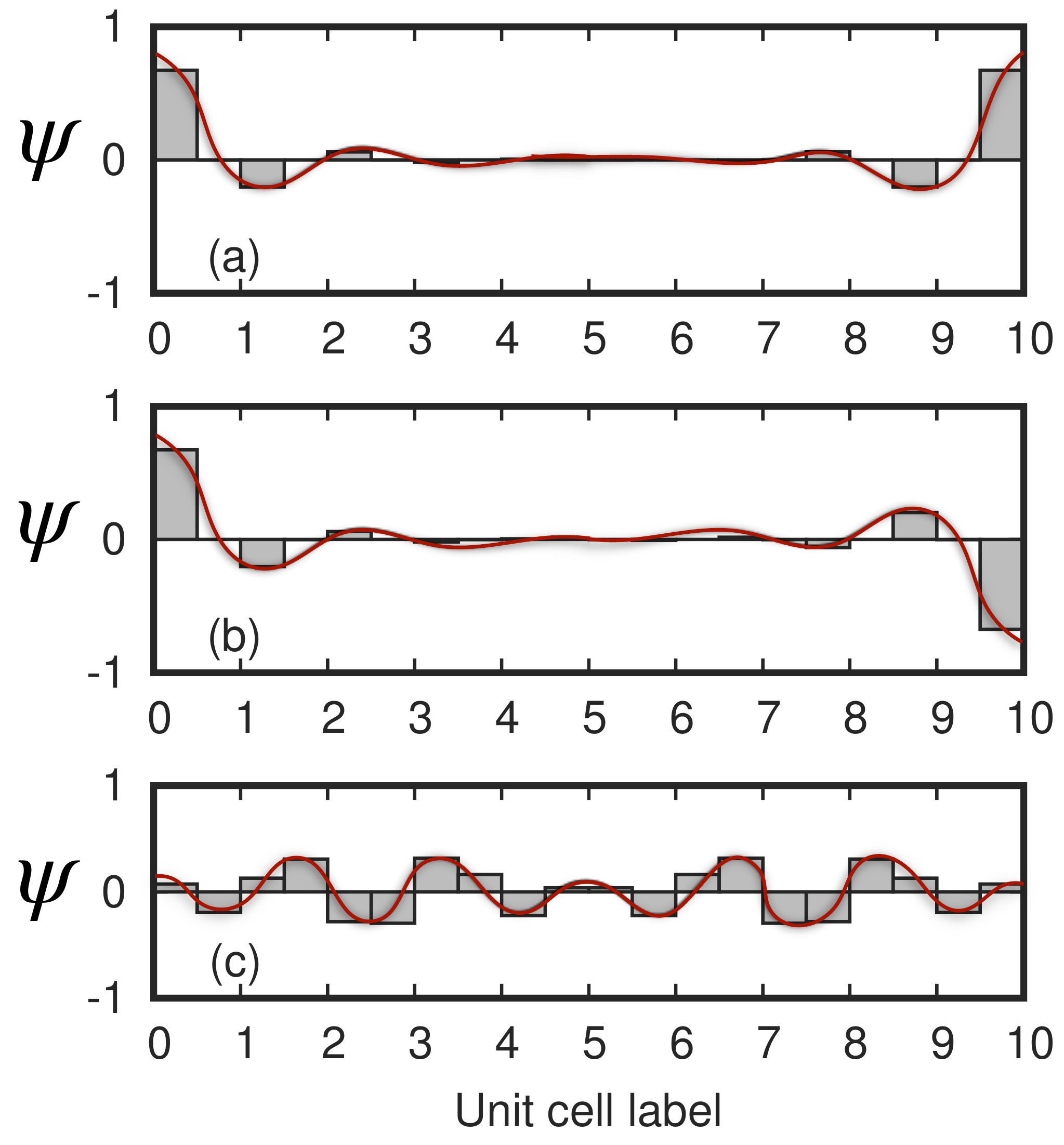}
\caption{Wavefunction amplitude at each site corresponding to (a) and (b) zero energy edge states and (c) non-zero energy eigenstate.}
\label{wavefunc}
\end{figure} 
Plots (a) and (b) in FIG. \ref{wavefunc} shows the (close to) zero energy states for the case when $v=0.3$ and $w=1.0$ and for reference plot (c), FIG. \ref{wavefunc} shows an arbitrary non zero energy state for the same case. We see that the non-zero energy state is delocalised throughout the chain. On the contrary, the zero energy states are exponentially localised at the edges of the chain which we expected from the dimerised limit. These states are called the \textbf{edge states} because they live on the edges of the chain. Also, notice that these edge state only remain zero energy edge states as long as $v<w$. Once $v>w$ there are no zero energy states and all states are delocalised throughout the lattice. The existence of these edge states in the $v<w$ case makes this case different from the case when $v>w$. Otherwise, in the bulk, for both these cases the system behaves like an insulator. 

We have found a physical consequence of the distinct bulk `topological' properties we found earlier - existence of Edge states. The case when the winding number is 1 is known as a topologically non-trivial case. Whereas the other is known as a trivial case. These two cases are related via a topological phase transition that corresponds to closing the band gap that gives rise to the conducting edge states. This is no coincidence! This is a simple example of the \textbf{bulk-boundary correspondence:} by looking at the bulk and calculating its topological invariants, in our case the winding number $\nu$, we can predict the existence of edge states at the boundaries or the other way around. 

\section{Conclusions}\label{sec:conclusions}
We started by solving the SSH model in the periodic boundary case which resulted in a bulk dispersion relation that is followed by the electrons. From there, we concluded that there are two different solutions: The metallic phase, when $v=w$ and the insulating phase, when $v\ne w$ which is possible when $w>v$ or $w<v$. From the bulk dispersion however, these two insulating phases seem to be equivalent. Then we argued that the information from the dispersion relation is incomplete! We then looked at an `abstract' quantity called the winding number that is governed by the eigenstates of the Hamiltonian in the bulk which attained different values for the two insulating cases. The winding number was then connected to the existence of the edges states at the boundaries of the model - this indeed proved that the two seemingly equivalent insulating phases are not quite equivalent at the boundaries. This non-trivial phase where there exists some boundary states and correspondingly a non-zero winding number in the bulk is known as a topologically non-trivial insulator whereas the other insulating phase which has zero winding number and correspondingly zero edge states, is a trivial insulator. The SSH model despite being simple captures all the essential features of topological insulators that are also encountered in higher dimensional topological condensed matter. 
 
\section{Acknowledgements} The idea of writing a review article on this topic originated during a discussion in the class of PHY665 (Quantum Phases of Matter and Phase Transitions) at IISER Mohali.


\end{document}